# The Concept of an Autonomic Avionics Platform and the Resulting Software Engineering Challenges

Bjoern Annighoefer
*University of Stuttgart*
*Institute of Aircraft Systems*
Stuttgart, Germany
0000-0002-1268-0862

Johannes Reinhart
*University of Stuttgart*
*Institute of Aircraft Systems*
Stuttgart, Germany
0000-0002-3512-5220

Matthias Brunner
*University of Stuttgart*
*Institute of Aircraft Systems*
Stuttgart, Germany
0000-0002-7941-2141

Bernd Schulz
*University of Stuttgart*
*Institute of Aircraft Systems*
Stuttgart, Germany
0000-0003-4538-600X

***Abstract*—The self-* properties commonly associated with the concept of autonomic computing are capabilities desirable for avionics software platforms. They decrease the configuration effort and inherently provide new fault tolerance and resource savings possibilities. The rigid certification process and the requirements for a static and predetermined system behavior are, however, in contradiction with the adaptive and flexible nature of autonomic computing systems. We propose a partition-based architecture providing autonomic features for avionics software platforms while being compliant to regulations and accepted technologies, such as ARINC 653. The core is a platform consciousness based on a domain-specific model and a novel MAP-QE-K cycle. Moreover, we suggest a planning intelligence, a virtual qualification authority, and a minimized execution unit. For each component we define the required design assurance level and possible realization techniques. We discuss the overall feasibility and point out central challenges in the fields of runtime verification and models at runtime. These challenges need to be solved up to the realization of autonomic avionics, e.g. a virtual security assessment and a qualifiable domain-specific model database.**

## I. INTRODUCTION

Modern aircraft are distributed computing platforms. The complexity and number of hosted system functions rises exponentially with each aircraft generation [1]. State of the art are generic avionics platforms that host several system functions in parallel by sharing the computing resources, e.g. Integrated Modular Avionics (IMA) [2]. Avionics platforms are assembled from a set of standardized avionics modules and bus systems and put into operation with configuration tables and software loads. Standardized avionics platforms reduce space, weight, volume and cost (SWAP-C) [3], but a huge effort exists for configuration, integration, and testing [4].

Within the project PAFA-ONE the technical feasibility of autonomic computing for safety-critical avionics platforms (according to CS-25) shall be shown as well as the potential path to certification. The concept is named Plug&Fly Avionics (PAFA). PAFA is an avionics platform that establishes its operational stage itself, while deriving configuration parameters, system function allocation, and fault tolerance on its own dynamically during runtime. This is completely different from current avionics systems, which are built and certified on the principle that every action is predetermined and verified during development; and that the system never leaves its statically defined configuration. This is a short position paper

- defining the PAFA concept, its usage, and applicable regulatory limitations;
- outlining a software architecture and consciousness model that potentially enables autonomic computing in the avionics world while being certifiable;
- introducing and discussing the major design decisions; and
- addressing important research challenges.

This paper is organized as follows. Chapter 2 discusses related works in the field of safety-critical autonomic computing systems. Chapter 3 defines the properties which PAFA shall exhibit and the limitations given by its application environment. Chapter 4 presents a partition-based architecture for PAFA. It introduces a planning intelligence with a novel MAP-QE-K cycle, a virtual qualification authority and a minimalistic execution unit. In Chapter 5 we present a domain-specific consciousness model. In Chapter 6 we propose a way forward in terms of research challenges that need to be addressed. The article concludes with Chapter 7.

## II. RELATED WORK

The research for adaptive avionics platforms started with the DIANA project [5] and was continued in the projects SYLVIA and KOMKAB. It resulted in a self-configuring cabin management system [6]. With the PAFA-ONE project this shall be extended to runtime adaptation, real-time capabilities, and greater topological flexibility. The baseline is autonomic computing [7]. We propose a PAFA-specific version of their Monitor-Analyze-Plan-Execute-Knowledge cycle (MAPE-K). The application of autonomic computing to cyber-physical and/or safety-critical and/or real-time systems, is addressed in [8], [9], [10], [11], [12], [13], [14], [15], [16] and [17]. In contrast to those works, PAFA targets primarily the configuration of the system. Compared to other self-* properties, self-configuration seems to be addressed less frequently within the autonomic community [18]. While not having the primary focus in our works, adaptivity of functions, environment interactions and fault tolerance can be beneficial for avionics systems. Moreover, our clear focus is the civil aircraft domain (CS-23/25) with the existing regulations and accepted standards enabling a certification. Besides that, we share the main concern of assurance, which is the evidence for the correct (and safe) functioning of a self-adaptive system, as summarized by de Lemos et al [19]. Within assurance our main issues are, however, not uncertainty, but runtime verification and models at runtime. Runtime verification is proposed by Tamura as checking self-adaptive functions against their requirements while the system operates [20].

While Tamura proposes runtime variants of technical verification methods, we address this by mimicking traditional aircraft certification activities during runtime, i.e. activities, that traditionally would have been carried out offline during development. That covers requirements verification as well as runtime safety and security assessments as for instance suggested in [21]. A foundation for our concept is the use of models at runtime as outlined in [22] and [23]. However, we do not want to run simulations of dynamic cyber-physical interaction, but we suggest a structural domain-specific model for function specifications and platform properties, containing all information required for our runtime verification. This is close to the problem of the verification of Dynamic Software Product Lines (DSPL), as proposed in [24] and [25]. It is, however, assumed that static configuration models and rules can be formally verified, which does not cover evidences on hardware compatibility or faulty reconfiguration. Our need is a certifiable runtime for such models. Last, we address the safe handover of one configuration to the next similar to Heinzemann [26], but with only one transaction.

## III. Plug&Fly Avionics and its Usage

A PAFA system is an autonomic computing system that is self-configuring, self-optimizing and self-healing. The information for taking decisions is maintained in a system-wide platform consciousness. The consciousness contains information on the tasks to be executed, the topology of the platform, the capabilities of the modules and networks, and the current function assignment. PAFA is a generic concept that is not bound to a specific hardware platform, e.g. safety-critical and real-time capable modules. Instead, the concept shall allow to specify features as real-time and hardware reliabilities such that they can be included in the platform decisions. The concept for PAFA is depicted in Fig. 1. A PAFA platform is a set of autonomic modules that are assembled to a PAFA system that is capable of hosting safety-critical functions. Therefore, it has to be capable of executing system functions and connect to system peripherals. The PAFA system establishes normal operation on its own, i.e. the topology of the PAFA system as well as the connected peripherals are detected automatically. Functions have to be provided to the PAFA system externally. The PAFA system decides on the current best assignment of tasks to the resources of the PAFA modules and network. Failure detection, reconfiguration, and degradation are conducted by the platform and are not part of the system function. The PAFA system determines the necessary degree of redundancy and generalizes failure detection and reaction mechanisms. It makes sure that the current execution instance of the system is compliant to the systems' performance and safety requirements. These requirements need to be supplied with the system logics. Therefore, a model-based function specification format is desired holding the logics and runtime requirements. The planning intelligence of the PAFA system is able to decompose the function specification in tasks (functional blocks), create the necessary redundancy patterns and map these to the hardware resources. This process is repeated if the system's topology or functions change, such that reconfiguration in case of failures or updates during maintenance are inherently supported by the platform.

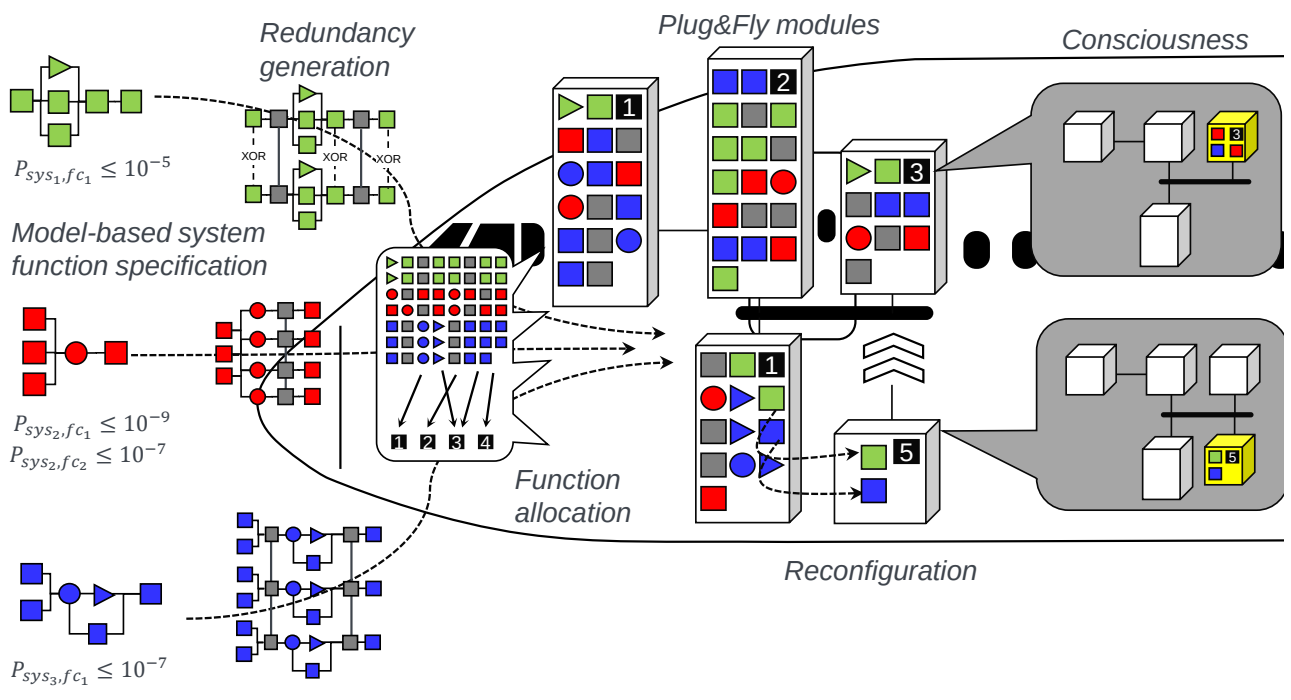

Fig. 1: Concept of Plug&Fly Avionics

### A. *System and Adaptation needs*

An avionics system is a service provider for aircraft system functions. Many of those system functions are safety-critical, i.e. humans die or are seriously injured if failures or malfunctions occur. Those typically include hard real-time control loops with time constraints of 1 ms and above. Three major sources for adaptations exist with different approximated occurrence frequencies: (1) hardware failures (~$10^{-4}$ 1/h); task/flight phase adaptation once every 10 min (~ 6 1/h) and (3) maintenance/upgrade on planned checks (~$10^{-2}$ 1/h). The configuration dimensions of a current aircraft are approximated by us with 4000 function and peripherals hosted on 50 generic avionics computers.

### B. *Relevant regulations and the certification process*

Aircraft regulations specify upper limits on the occurrence probabilities of failure conditions caused by systems. For instance, it is mandatory that catastrophic failure conditions for systems of large civil aircraft have a maximum occurrence probability of $10^{-9}$ per flight hour and cannot be caused by a single failure (cf. CS-25 1309 and AMC 1309). Considering realistic hardware failure rates of approximately $10^{-5}$ per hour, it is clear that this is not solved by selecting the appropriate hardware components, but by developing an appropriate redundancy structure. The system architecture contributes essentially to the required failure probabilities by having sufficient replicas. Since empirical verifications of failure rates within the required orders of magnitude are infeasible, the safety in avionics is based on a rigid and audited development process (ARP 4754) accompanied by safety (ARP 4761) and security (DO-326) assessment methods. In summary, used hardware, topologies, and fault tolerance mechanisms must be shown to have a probability of relevant failure conditions below defined limits. Moreover, hardware and software must be developed according to strict validation and verification plans (RCTA DO-254 and DO-178C). Based on the criticality, Design Assurance Levels (DAL) E (low criticality) to A (high criticality) are assigned. DALs define mandatory requirements and test coverages. Finally, certification is only granted for the final vehicle, not for a piece of hardware or software. Avionics platforms host system functions in parallel. Certification is only issued for the fully integrated avionics system. However, standards have been established enabling the provision of certification credits by the platform [27]. Not all tests have to be carried out for the integrated system as long as integrated systems are compliant to the usage rules of the platform.

### C. *Current avionics systems and platforms*

Most popular is the approach of Integrated Modular Avionics (IMA). IMA implements a strict temporal and spatial partitioning for computing and communication. Partitioning ensures that system functions cannot influence each other if not intended. The hardware and operating system of IMA modules must guarantee partitioning under normal and failure conditions. System function can be developed and integrated independently (cf. DO-297). ARINC 653 is the accepted standard for providing partitioning on the operating system level. System functions are developed on top of an ARINC 653 API. ARINC 653 assigns fixed memory regions and fixed periodic timeslots to system function. Moreover, it defines a unified interface to bus system and I/O interfaces. The real hardware is transparent for the system function. The properties of partitions and the connection to I/Os is defined by configuration parameters that can only be changed on load time and are static during runtime. All system specific failure detection and fault tolerance mechanism are out-of-scope of the IMA system and are part of the system software loaded to the IMA modules. Consequently safety-critical IMA system partitions exist on several IMA modules in parallel. The fault tolerance mechanisms are system-specific and need to be proven as part of the system development process.

## IV. A SAFETY-CRITICAL PARTITION-BASED ARCHITECTURE

In order to be as close as possible to the regulation and accepted standards of certified civil aircraft systems, we propose an architecture for autonomic computing that is based on strict partitioning and can reside above an ARINC 653 API. This partition-based autonomic computing architecture is outlined in Fig. 2. The basic principle is that of a Chimera partition, i.e. a statically configured partition with fixed memory and partition time window, but variable content. The Chimera partition holds a simple function execution unit, which executes logics of several system functions as desired by the current active configuration. The execution schedule and signal flow are determined by a planning intelligence within a separate partition. Before a configuration is forwarded to the execution unit, it is verified by a virtual qualification unit, which comes together with a verification execution partition. Not all qualification activities are virtualized. Like for IMA, the PAFA platform and system functions must qualify aspects that are independent of the configuration beforehand, i.e. there will be a human designated engineering representative (DER) at the platform provider, platform integrator, and system developer. All partitions are situated on a partitioned (real-time) operating system, e.g. with an ARINC 653 API. Moreover, a signal-based communication abstraction relieves system functions from knowing the target device and physical connections. In the following, the purpose and the functional principle of the four partitions is further detailed.

### A. *Live Execution Unit*

There are two execution unit partitions. The live execution unit is responsible for running system functions. The system functions are composed of smaller building blocks, so-called basic tasks. The execution unit gets a task schedule for each system function or part of a system function. See Fig. 3 for an example of a system function composed of basic tasks and Fig. 4 for the resulting schedule. The schedule must be in line with the partition window of the live execution partition. It is assumed that system functions are either periodic and guaranteed to be executed within a known worst case time, or they are best-effort non-periodic tasks. Moreover, the execution unit hosts a routing table that maps signals between basic tasks either to internal memory buffers or external communications. The execution unit is safety-critical since it must be able to execute system functions of any criticality. Therefore, it must be implemented in accordance to DAL A. Consequently, it is best to keep the internal logic as minimalistic as possible. The execution unit does not take any decisions and runs the tasks and signal configuration until reconfiguration. The schedule and the routing table are assumed to be correct before entering the execution unit. A partitioning of system functions is not necessary, because they are composed from basic building blocks that are developed in accordance to DAL A and have to be proven to have no side-effects. In addition, the composition of blocks has to be proven to be correct, e.g. by a contract-based formal verification approach carried out beforehand offline, i.e. defined formal pre and post conditions for each task that can be checked by a theorem proofer to be not violated by the system specification and resulting in the desired behavior.

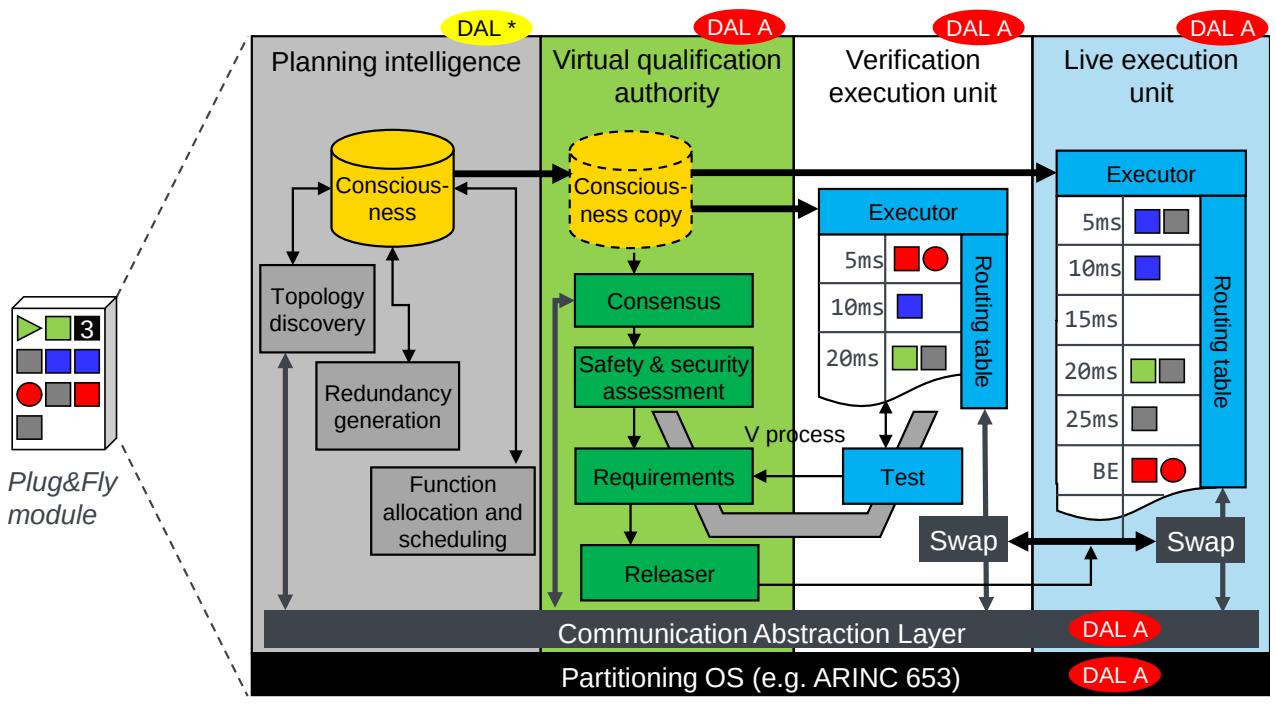

Fig. 2: A partition-based architecture of a Plug&Fly avionics module

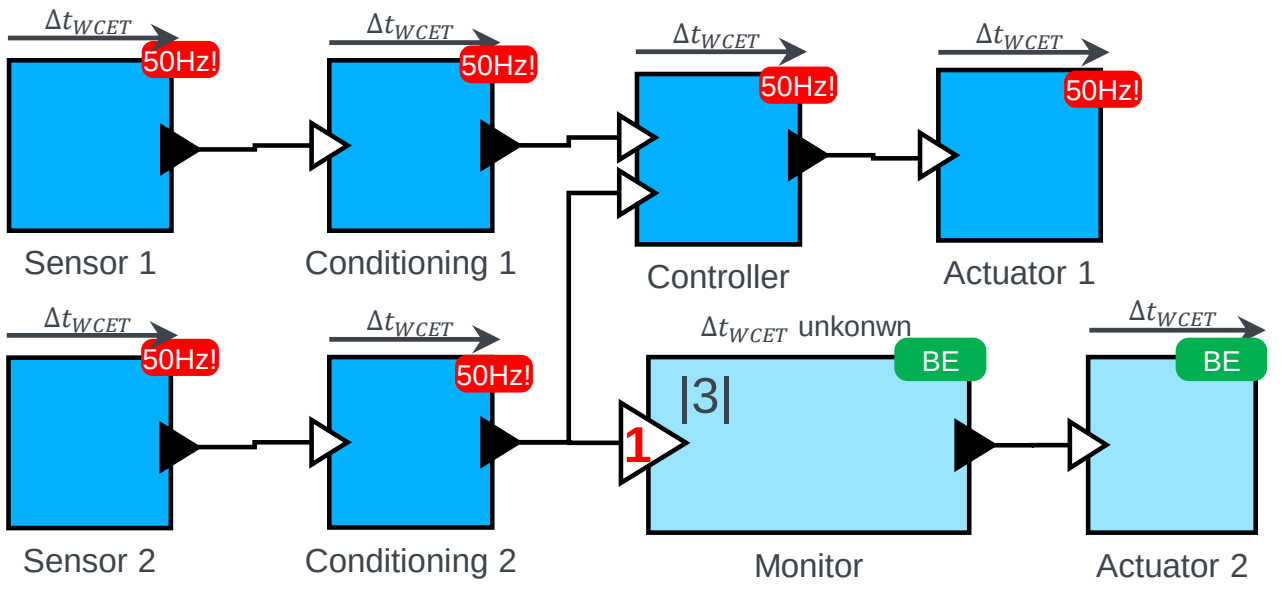

Fig. 3: Example of a system function composed of basic tasks, a controller architecture with two sensors,one actuator, and monitor (BE = best effort, WCET = worst-case execution time)

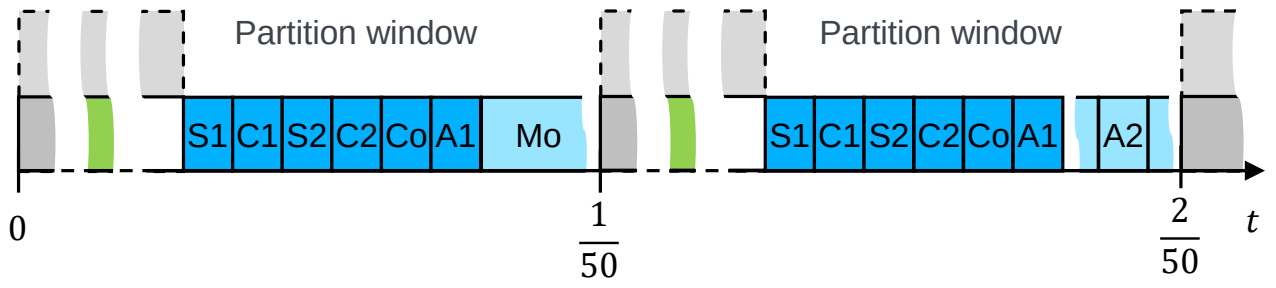

Fig. 4: Schedule for basic tasks on one module within partition windows.

### B. *Planning Intelligence*

The planning intelligence comprises monitor, analyze, and plan of a MAPE-K loop. It is responsible for detecting changes in topology, device status, resource availability, and demanded system functions. The information retrieved by topology discovery and user inputs is managed within the platform consciousness. The consciousness contains the information on all PAFA modules participating in the system, the desired function specifications, and the currently executed configuration. Based on the consciousness, the planning intelligence derives the current best-effort configuration and decides if adaptation is necessary. Best-effort means, the platform executes all system functions with the required performance and safety levels. If, however, that is infeasible with the available resources, a degraded configuration is determined, which prioritizes functions and reduces safety levels as necessary. The planning process starts with the generation of redundancies, i.e. the basic tasks of the system function are replicated in accordance to its failure conditions and the properties of the hardware. In addition, fault detection and tolerance mechanisms are inserted, i.e. reliable broadcast and voting. Afterwards, all basic task instances are assigned to modules while considering the available resources, segregation rules, execution times, and communication delays. This is a function allocation and scheduling problem, which could be solved by combinatorial optimization or search-based solutions. The schedule is the input for the execution unit. Map, analyze, and plan activities are complex operations. It is unlikely that in near future deterministic, fast and simple algorithms are available, such that the planning intelligence partition can be qualified, i.e. a high DAL is unlikely. Therefore, the plan is not forwarded to the execution unit, but verified by a virtual qualification authority. It is not assumed that the required safety level is achieved by adaptation, but with traditional redundancy mechanisms. This also increases the time for planning, because in case of a failure no immediate reaction is required. However, with adaptation, availability can be advanced over the required level.

### C. *Virtual Qualification Authority*

The virtual qualification authority is the gatekeeper for the execution unit. Before a new schedule and routing table (the so-called plan) becomes active, it is verified. This is necessary, firstly because the planning intelligence is assumed to be low DAL and secondly because current software certification assumes that each configuration is covered by certification. It must also be robust against faulty low-DAL input. Having all valid configuration options qualified beforehand (cf. [28], [29], [30], [31]) is in principle possible, but restricts the autonomic capability. Therefore, we propose to mimic activities of a qualification process by the platform itself within a virtual qualification authority partition. A qualification activity is the demonstration of compliance to the objectives of ARP4754, DO178C, ARP4761, etc. While most objectives can be addressed with formal (mathematical) methods (cf. DO-333), a few remain as explicit manual review and validation activities as well as ones requiring independent compliance evidences. Today, qualification authorities expect documents as evidences. A virtual qualification authority is incompatible with current certifications procedures, but we think an equivalent level of safety can be reached. The virtual qualification activities are as follows:

1) It has to be verified, that the decision base for the configuration change is correct and unambiguous. Therefore, a verification of identified topologies takes place with fraud-resistant identification checks, e.g. by digital signatures or blockchains. Moreover, the plan has to be unique within the PAFA system. Therefore, a consensus procedure is executed between the virtual qualification authorities of the PAFA modules. This happens in the consensus block.

2) After a unique plan exists in all virtual qualification authority instances, the required safety and security have to be verified by autonomic safety and security assessments. The module properties (e.g. failure mode occurrence rates) and the proposed schedule and routing are used to calculate the probabilities of failure conditions and threat risks. These are then compared to the system specifications. In case of safety, the automatic derivation of fault trees is suggested. In case of security the generation of attack trees is a promising solution. Both require that failure and threat propagation information is part of the consciousness (s. next chapter). The derived probabilities must be below the boundaries specified for the systems.

3) The traditional development process follows a requirements-based V-process. A real certification authority supervises the process and spot checks requirements, implementations, test cases, test reports, and their traceability. The verification of implementations is out of scope for the virtual qualification authority since only qualified basic tasks are concatenated. The correct execution of the system function has, however, to be verified as well as the required performance (e.g. latencies and synchronicity). It is assumed that the system specification is provided as a model (s. next chapter), i.e. the planning intelligence's plan is verified against model-based requirements not textual ones (which is accepted according to DO-331). As such, it must be verified that a plan results in the system functions executing as desired by their specifications. This is supposed to be carried out by a passive verification execution unit, which executes the new schedule, but has no interaction with actual system operation, e.g. actuator signals are not forwarded. Since this is not sufficient to verify real world interactions, missing signals in the verification execution need to be augmented with signals from the live execution unit or recorded system behavior. A validation of system functions can be carried out offline and is, therefore, not part of the virtual qualification authority.

4) If virtual verification activities have a positive outcome, the plan is cleared to be executed. This decision is made consistent between replicas with another consensus exchange. The decision is forwarded to the execution unit. In order to make the transition process from the current to the new system configuration a short and deterministic transaction, the live execution and verification execution unit can swap roles immediately by changing their routing tables simultaneously.

With this, the autonomic cycle is extended to a redundant MAPE-QE-K cycle as depicted in Fig. 5.

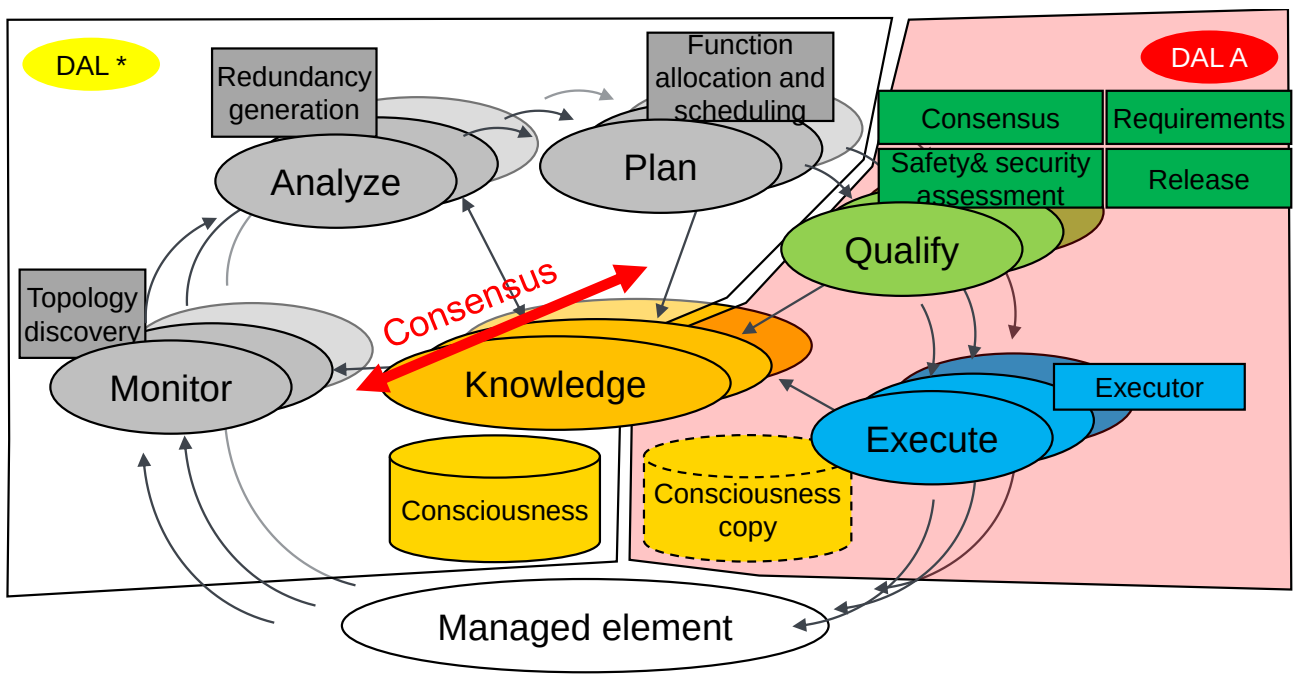

Fig. 5 A redundant MAP-QE-K circle with Design Assurance Levels (DAL).

Q denotes the virtual qualification activity as mentioned above. Q and E are separated from the rest, since they require a higher design assurance level. For the virtual qualification authority to be accepted as an adequate replacement of parts of the human qualification process, it is necessary that all verification artifacts are accessible and explainable on demand. Since the virtual qualification authority has the control on the system function execution, it needs to be developed in accordance to DAL A, which is a huge challenge.

If no plan exists or is rejected, no adaptation takes place. This does not cause an unavailability of the functions, since traditional redundancies are in place. To recover there are two options: (1) try to find a plan with degraded system functions. (2) try to find a best-effort solution with deactivated redundancies and functions based on their priority and indicate maintenance needs.

### D. *Verification execution unit*

The verification execution unit is technically identical to the live execution unit. It executes a schedule of basic tasks and routes the signals internally or to the communication abstraction according to a routing table. Its purpose is to test future adaptations in the real environment, while having no influence on the current operation of the PAFA system. Therefore, all outgoing signals are passivated. A communication between the verification partitions of multiple PAFA modules is possible. This is realized by a modified routing table configuration provided by the virtual qualification authority. The routing table has a switch input reserved for the virtual qualification authority, such that within one cycle the verification unit can be set to the live execution unit and the live execution unit becomes the verification execution unit. Switches are coordinated with consensus algorithms. If the test execution or switching fails, replanning and traditional redundancy are in charge.

## V. A DOMAIN-SPECIFIC MODEL AS THE CONSCIOUSNESS

A domain-specific consciousness takes the role of the knowledge (K) in the MAPE-QE-K cycle of the PAFA system. We name it consciousness since this represents the fact that it holds the self-aware platform state with all the information required to take reliable self-organization decisions. Domain-specific technology was chosen since it enables the representation of complex data and relationships and ease machined processing and verification, i.e. a domain specific model has a rigid structure based on a small set of basic elements, easy to check. Moreover, formal constraints can be attached to domain-specific elements. There are two consciousness instances: (1) A consciousness model updated continuously by monitor and planning. (2) A verified consciousness resulting from a consciousness copy forwarded from planning to the virtual qualification partition when a new plan is necessary. The consciousness model contains

- the topology of participating PAFA modules and peripherals, i.e. the physical connection, interfaces, and physical location information (locations are mandatory for some system function);
- the device's resources and functional capabilities, i.e. available resources, available basic tasks, resource needs per task, and worst-case execution times;
- device failure modes and probabilities;
- the function logics
- the function safety, security, and performance requirements, i.e. failure conditions, threat conditions, and maximum occurrence probabilities as well as periods, latency, and synchronicity settings;
- the current allocation of tasks to devices and resources and schedules (current PAFA configuration); and
- the current allocation of signals to devices or connections, schedules, and resources.

As a prototype, the Open Avionics Architecture Model (OAAM, www.oaam.de) was created. OAAM is a model based on the ECORE meta-modeling language of the Eclipse Modeling Framework. Its meta-model is documented in [32]. OAAM is available as open source and is actively used in the planning and automatic instantiation of avionics systems such as IMA [33]. Therefore, we are certain, that it includes already most of the parameters necessary for a self-configuring avionics platform.

OAAM comprises the layers library, scenario, systems, functions, hardware, anatomy, capabilities, restrictions, and allocations. The central elements are resources that can be defined in arbitrary types. Resources are provided by devices in certain numbers and consumed by the allocation of tasks and signals. Capability objects specify the exact relation between task/signal types, capable device/connection types, and consumed resources. Moreover, capabilities define worst-case execution/transmission times. The hardware layer contains topologies composed of devices, interfaces and connections. All hardware objects are of a certain type defining their resources and properties, such as failure modes and probabilities. If necessary, those can be refined based on the installation location, which are modeled in the anatomy layer (e.g. to consider different vibration, temperature, radiation or humidity levels). The functions layer is used to express the function specification as a composition of basic tasks connected by signals (cf. Fig. 3). Moreover, it contains execution rates, peripherals, latency and synchronicity demands, failure conditions, and safety requirements. It is also possible to specify functions that scale with the topology, parallel executions, tasks of unknown length, and variants. Please find the details on possible function-specification in [34]. The assignment of tasks to devices and signals to connections or devices is expressed in the allocations layer with assignment objects. An assignment object defines the task/signal and device/connections object as well as the capability used. Moreover, a schedule can be attached. This represents a configuration of the PAFA system. All layers in OAAM can exist multiple times, such that the current configuration and arbitrary future configurations can exist in the same model without the necessity to replicate any of the basic objects. Currently not included in OAAM are final failure and threat propagation models, security requirements, degraded modes, and system priorities. Those extensions are under development and will be added in the near future.

## VI. CHALLENGES RESULTING FROM PLUG&FLY AVIONICS

We are certain, that the proposed architecture and model are technically feasible and that in the future, a qualification of the PAFA platform can be possible. However, certification will require that several research challenges are addressed beforehand. One challenge is the development of the planning algorithm that takes all required information into account and comes up with a meaningful best effort solution in an acceptable time. This should be possible with the combination of available combinatorial optimization approaches. Another challenge is the fraud-resistant detection and especially localization of devices and peripherals. It is assumed that methods from zero-trust as well as trusted computing can be utilized to close that gap. We see major challenges in using models at runtime and runtime verification.

PAFA requires the consciousness model at runtime. Since the consciousness is part of the virtual verification authority partition, its implementation is required to be qualified in accordance to DAL A. Such a runtime is not available. Current domain-specific model runtimes have different scopes and are implemented in languages that makes the handling of object-oriented data easy, e.g. Java or Python. This is not compatible to aircraft certification. It would be possible to generate a target-specific implementation in a certification-compatible language, such as C or ADA. However, that would either require a qualified model generator (not available) or the manual qualification of a huge code-base (OAAM has ~190 classes) and requalification after each change. It would be more efficient to have a minimalistic runtime that solely implements the meta-meta-model. A meta-meta-model (e.g. EMOF) usually has only a low number of essential classes (<10). Domain-specific models are then interpreted during runtime. In that way, all benefits (e.g. reflections) of meta-modelling would be available while having minimal code footprint. However, it is still unclear, which features of meta-meta-models will have to be removed in order to ensure deterministic behavior, e.g. opposite references, event handling, unlimited recursions or multiplicities. Moreover, it is unclear, which restrictions or proofs will have to be applied to user models and connected algorithms in order to keep up the trustworthiness of the runtime. Another challenge is the machined processing of the consciousness in algorithms. Using a conventional programming language API for model access is not very efficient and produces a large code footprint with many repeating actions. That is again hampering certification. For instance, model search and filter operations are frequently necessary and should be implemented generically. It is suggested to provide model access and modification operations in a more high-level model query language, but with deterministic behavior, such that the interfacing code of algorithms becomes small and the query engine is still simple enough to be qualified. Essential Object Query (EOQ, gitlab.com/eoq/essentialobjectquery) is a first prototype of what is suggested. We investigate a qualified meta-meta-model runtime and the query language implementation within the TALIA project.

The purpose of the virtual qualification authority is the runtime verification of the PAFA system before each configuration change. The proposed way is to mimic manual qualification activities. Challenge one is to find algorithms that are small and deterministic to carry out online tests, safety assessments, and security assessments. While test generation and automatic safety artifact generations are well understood research topics with formalized approaches, automatic security assessments are quite virgin. Concerning the security assessment, fault-propagation models exist (also inside OAAM) with which an automatic derivation and evaluation of fault trees is possible. Minor work remains to include external events as well as automatic common cause and risk analyses. Considering security, the methods proposed in aircraft certification are similar to safety, e.g. threat trees analysis. The relation of threat propagation in the avionics system, however, and the modeling of threat propagation is not well-formalized. Current automatic approaches require still a lot of manual inputs to derive threat trees [35]. Moreover, formal methods are fine as long as their results can be reviewed by a human. Since that is not possible the implementation of the formal method itself needs qualification (cf. DO-330). An open issue is also the compliance to objectives requiring independent evidences. It might be possible to drop independence if the methods implementation is qualified. Otherwise, two dissimilar virtual qualification authorities would be needed.

A further challenge is the consensus of complex model data. In principle traditional consensus routines can be applied on individual model elements, i.e. reliable broadcast with voting. However, with increasing model size and many PAFA modules, this becomes infeasible. More lightweight approaches are required, e.g. a hierarchical consensus. Finally, solving runtime qualification algorithmically is not sufficient. In order to be accepted by human qualification authorities, the decisions taken have to be understandable and traceable. Therefore, equivalents to the paper-based certification artifacts have to be found. This will be a matter of on-demand artifact generation and an appropriate visualization/GUI of system states and decisions.

We will work further on all the challenges while the PAFA-ONE project proceeds.

## VII. Conclusion

Plug&Fly Avionics (PAFA) can decreases development efforts for avionics systems significantly. PAFA is a real-time capable and safety-critical autonomic computing platform that is self-configuring, self-optimizing, and self-healing. It finds its best- effort operational state for a given set of system functions with additional performance and safety requirements. A practical realization has to be compliant with accepted avionics standards, such as strict partitioning and general aircraft certification regulations. We propose an architecture that comprises four static partitions on each PAFA module: a live execution unit, a planning intelligence, a virtual qualification authority and a passive verification execution unit. This setup realizes a MAP-QE-K cycle that isolates complex planning algorithms (MAP) in a partition that requires a low Design Assurance Level. The remaining partitions (Q and E) require DAL A, but are much simpler. The aircraft operating system API ARINC 653 can be reused. We suggest to use the open source domain-specific model OAAM as the knowledge base for autonomic decisions. OAAM contains a resource model that enables a fine-grained modeling of task and signal assignment within a generic hardware topology. It provides elements to model performance and safety requirements, fault propagations, timings, and hardware failure modes. The main challenges on the way to the realization of PAFA are first a qualifiable meta-meta-model runtime with an efficient model query interface and second a runtime verification that mimics traditional certification activities at least equivalent to human auditing capabilities. Requirements-based testing as well as safety and security assessment have to be carried out in a qualifiable manner, which requires deterministic implementations and a self-explaining interface. Further progress will be made within the research project PAFA-ONE in cooperation with the European Union Aviation Safety Agency (EASA).

## Acknowledgment

This paper is based on the PAFA-ONE project (contract: 20E1921) funded by the German Federal Ministry of Economic Affairs and Energy (BMWI) in the LuFo VI-1 program.

## References

[1] J.-B. Itier, *IMA1G - Genesis and Results,* 2009.

[2] P. J. Prisaznuk, "Integrated Modular Avionics," in *NAECON*, 1992.

[3] A. Mairaj and R. Tahir, "SWaP Reduction: Vital for Choice of Avionics Architecture," in *International Conference on Engineering and Emerging Technologies, Lahore, Pakaistan*, 2014.

[4] M. Halle and F. Thielecke, "Next generation IMA configuration engineering - from architecture to application," in *DASC*, 2015.

[5] R. Reichel, R. Ahmadi and M. Lehmann, "Self-Adaptive Avionics Platform," in *AST 2011*, 2011.

[6] B. Annighoefer, M. Riedlinger, O. Marquardt, R. Ahmadi, B. Schulz, M. Brunner and R. Reichel, "The Adaptive Avionics Platform," *IEEE Aircraft Electronics Systems Magazine,* 2019.

[7] J. O. Kephart and D. M. Chess, "The vision of autonomic computing," *Computer,* vol. 36, pp. 41-50, 1 2003.

[8] E. T. McGee and J. D. McGregor, "Using Dynamic Adaptive Systems in Safety-Critical Domains," in *SEAMS*, 2016.

[9] A. Petrovska, S. Quijano, I. Gerostathopoulos and A. Pretschner, "Knowledge aggregation with subjective logic in multi-agent self-adaptive cyber-physical systems," in *SEAMS*, 2020.

[10] A. Bennaceur, C. Ghezzi, K. Tei, T. Kehrer, D. Weyns, R. Calinescu, S. Dustdar, Z. Hu, S. Honiden, F. Ishikawa, Z. Jin, J. Kramer, M. Litoiu, M. Loreti, G. Moreno, H. Muller, L. Nenzi, B. Nuseibeh, L. Pasquale, W. Reisig, H. Schmidt, C. Tsigkanos and H. Zhao, "Modelling and Analysing Resilient Cyber-Physical Systems," in *SEAMS*, 2019.

[11] G. Moreno, C. Kinneer, A. Pandey and D. Garlan, "DARTSim: An Exemplar for Evaluation and Comparison of Self-Adaptation Approaches for Smart Cyber-Physical Systems," in *SEAMS*, 2019.

[12] S. Bhattacharyya, D. Cofer, D. Musliner, J. Mueller and E. Engstrom, "Certification considerations for adaptive systems," in *ICUAS*, 2015.

[13] G. Weiss, P. Schleiss, D. Schneider and M. Trapp, "Towards integrating undependable self-adaptive systems in safety-critical environments," in *SEAMS*, 2018.

[14] M. Camilli, A. Gargantini and P. Scandurra, "Specifying and verifying real-time self-adaptive systems," in *ISSRE*, 2015.

[15] A. Borda, L. Pasquale, V. Koutavas and B. Nuseibeh, "Compositional Verification of Self-Adaptive Cyber-Physical Systems," in *SEAMS*, 2018.

[16] R. Calinescu, D. Weyns, S. Gerasimou, M. U. Iftikhar, I. Habli and T. Kelly, "Engineering Trustworthy Self-Adaptive Software with Dynamic Assurance Cases," *IEEE Transactions on Software Engineering,* vol. 44, pp. 1039-1069, 2018.

[17] T. Kluge, "A role-based architecture for self-adaptive cyber-physical systems," in *SEAMS*, 2020.

[18] J. P. S. da Silva, M. Ecar, M. S. Pimenta, G. T. A. Guedes, L. P. Franz and L. Marchezan, "A systematic literature review of UML-based domain-specific modeling languages for self-adaptive systems," in *SEAMS*, 2018.

[19] R. Lemos, D. Garlan, C. Ghezzi, H. Giese, J. Andersson, M. Litoiu, B. Schmerl, D. Weyns, L. Baresi, N. Bencomo, Y. Brun, J. Cámara, R. Calinescu, M. Cohen, A. Gorla, V. Grassi, L. Grunske, P. Inverardi, J.-M. Jézéquel and F. Zambonelli, "Software Engineering for Self-Adaptive Systems: Research Challenges in the Provision of Assurances," 2017, pp. 1-29.

[20] G. Tamura, N. M. Villegas, H. A. Müller, J. P. Sousa, B. Becker, G. Karsai, S. Mankovskii, M. Pezzè, W. Schäfer, L. Tahvildari and K. Wong, "Towards Practical Runtime Verification and Validation of Self-Adaptive Software Systems," in *Software Engineering for Self-Adaptive Systems II: International Seminar,* 2013, p. 108–132.

[21] S. Kabir, I. Sorokos, K. Aslansefat, Y. Papadopoulos, Y. Gheraibia, J. Reich, M. Saimler and R. Wei, "A Runtime Safety Analysis Concept for Open Adaptive Systems," in *Model-Based Safety and Assessment*, Cham, 2019.

[22] B. H. C. Cheng, K. I. Eder, M. Gogolla, L. Grunske, M. Litoiu, H. A. Müller, P. Pelliccione, A. Perini, N. A. Qureshi, B. Rumpe, D. Schneider, F. Trollmann and N. M. Villegas, "Using Models at Runtime to Address Assurance for Self-Adaptive Systems," in *Models@run.time*, 2014, p. 101–136.

[23] T. Vogel, "mRUBiS: An Exemplar for Model-Based Architectural Self-Healing and Self-Optimization," in *SEAMS*, 2018.

[24] I. S. Santos, L. S. Rocha, P. A. S. Neto and R. M. C. Andrade, "Model Verification of Dynamic Software Product Lines," in *Proceedings of the 30th Brazilian Symposium on Software Engineering*, 2016.

[25] R. Olaechea, J. Atlee, A. Legay and U. Fahrenberg, "Trace Checking for Dynamic Software Product Lines," in *SEAMS*, 2018.

[26] C. Heinzemann, S. Becker and A. Volk, "Transactional execution of hierarchical reconfigurations in cyber-physical systems," *Software & Systems Modeling,* vol. 18, p. 157–189, 2 2017.

[27] A. Wilson and T. Preyssler, "Incremental Certification and Integrated Modular Avionics," in *DASC*, 2008.

[28] Z. Li, S. Wang and T. Zhao, "A model based simulation verification method for IMA reconfiguration on system level," in *2015 First ICRSE*, 2015.

[29] D. Suo, J. An and J. Zhu, "A new approach to improve safety of reconfiguration in Integrated Modular Avionics," in *DASC*, 2011.

[30] G. Montano and J. McDermid, "Human involvement in dynamic reconfiguration of Integrated Modular Avionics," in *DASC*, 2008.

[31] Q. Zhang, S. Wang and B. Liu, "Approach for integrated modular avionics reconfiguration modelling and reliability analysis based on AADL," *IET Software,* vol. 10, pp. 18-25, 2016.

[32] B. Annighoefer, "An Open Source Domain-specific Avionics System Architecture Model for the Design Phase and Self-organizing Avionics," in *SAE AeroTech Americas*, 2019.

[33] B. Annighoefer, M. Brunner, J. Schoepf, B. Luettig, M. Merckling and P. Mueller, "Holistic IMA Platform Configuration using Web-technologies and a Domain-specific Model Query Language," in *DASC*, 2020.

[34] B. Annighöfer, M. Riedlinger and O. Marquardt, "How to tell configuration-free integrated modular avionics what to do?!," in *DASC*, 2017.

[35] B. Solhaug and K. Stølen, "The CORAS Language – Why It Is Designed the Way It Is.," in *Safety, Reliability, Risk and Life-Cycle Performance ofStructures & Infrastructures*, 2013.